\documentclass[12pt]{iopart}
\usepackage{iopams}
\usepackage{graphicx}

\begin{document}

\title[QCD sum rules for $D$ mesons in dense and hot nuclear matter]{QCD sum rules for $\bf D$ mesons in dense and hot nuclear matter}

\author{T. Hilger, R. Schulze and B. K\"ampfer}

\address{Forschungszentrum Dresden-Rossendorf, PF 510119, D-01314 Dresden, Germany}
\address{TU Dresden, Institut f\"ur Theoretische Physik, D-01062 Dresden, Germany}

\ead{t.hilger@fzd.de}
\begin{abstract}
Open charm mesons (pseudo-scalar and scalar as well as axial-vector and vector)
propagating or resting in nuclear matter display an enhanced sensitivity to the chiral
condensate. This offers new prospects to seek for signals of chiral restoration, in particular in $pA$ and $\bar{p}A$ reactions as envisaged in first-round experiments by the CBM and
PANDA collaborations at FAIR. Weinberg type sum rules for charming chiral partners
are presented, and the distinct in-medium modifications of open-charm mesons
are discussed. We also address the gluon condensates near $T_c$ and their impact on QCD sum rules.
\end{abstract}

\submitto{\JPG}
\maketitle

\section{Introduction}

The chiral condensate seems to be of utmost importance as candidate of an order parameter of chiral symmetry which is spontaneously and explicitly broken in nature (vacuum).
The approach to restored chiral symmetry in a dense and/or hot medium may be signaled by a modified hadron spectrum.
In particular, the spectral strengths of hadrons are expected to become modified when embedded in strongly interacting matter.
While for vector mesons and strange mesons fairly exhaustive studies have been performed in this context, charm degrees of freedom require further investigations.
The charmed hadron spectroscopy in $pA$ and $\bar{p} A$ collisions will be addressed by two major experimental set-up's of FAIR, CBM and PANDA.
With this motivation we report in section \ref{sct2} our results of a QCD sum rule analysis of spectral changes of pseudo-scalar open charm $D$ mesons in cold nuclear matter.

In section \ref{sct3} we consider differences of spectral strengths of chiral partners thus exposing the role of the chiral condensate also for charm.
Since, within the QCD sum rule approach which we employ here, the temperature and density dependence of QCD condensates is important, we present in section \ref{sct4} a first explorative study of two gluon condensates near the deconfinement temperature but at finite net baryon density.

\section{QCD sum rules for $\bf D$ and $\bf \bar{D}$ mesons}
\label{sct2}

\begin{figure}[ht]
\centering
\includegraphics[width=0.3\textwidth]{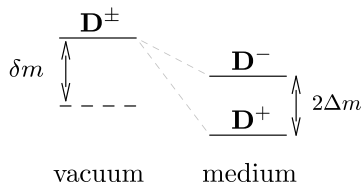}
\caption{Expected spectral changes of $D$ mesons in a cold nuclear medium.}
\label{fig1}
\end{figure}

\begin{figure}[ht]
\centering
\includegraphics[width=0.45\linewidth]{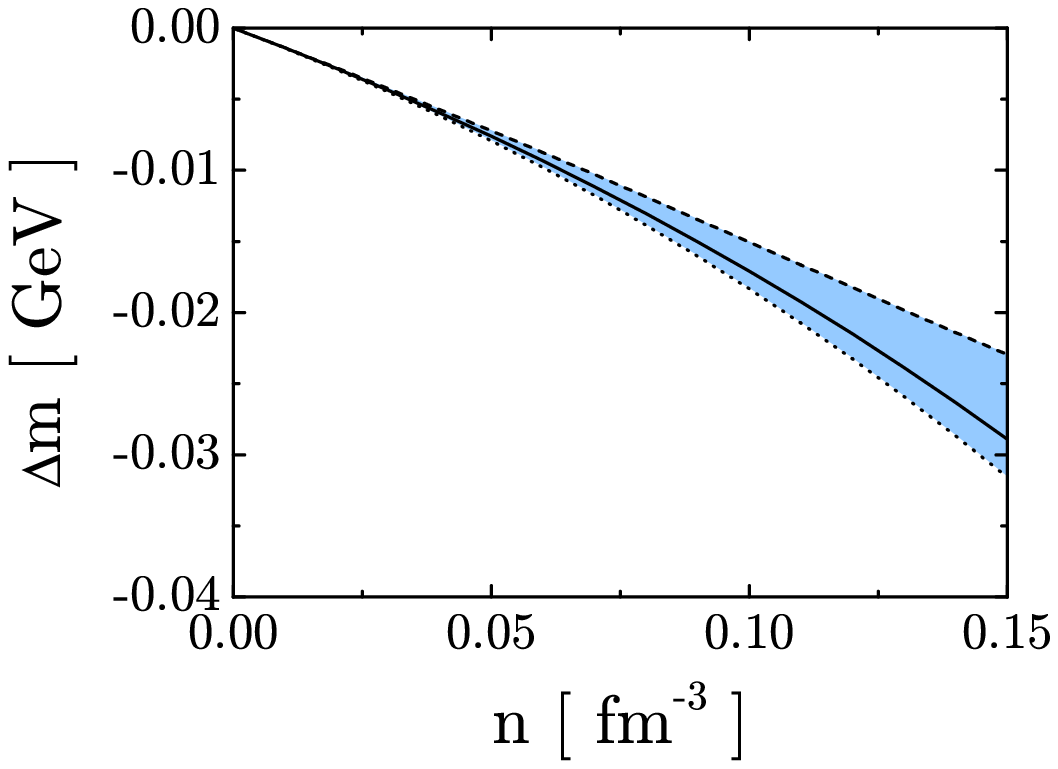}
\includegraphics[width=0.45\linewidth]{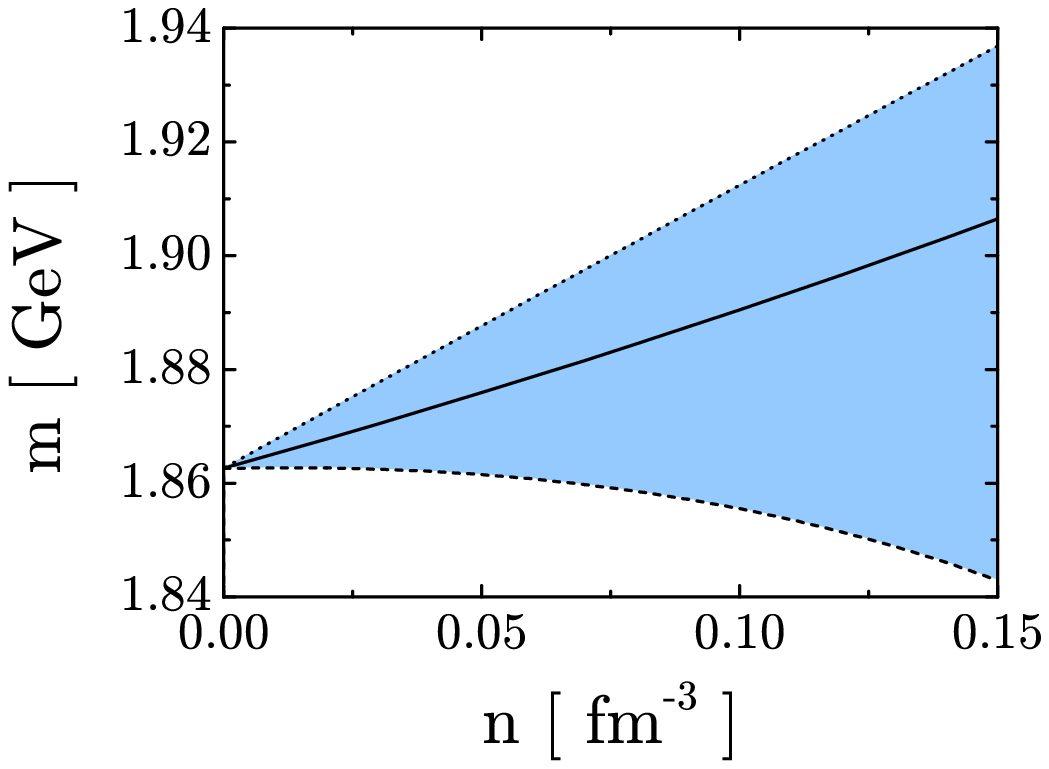}
\caption{(color online) The $D$-$\bar{D}$ mass splitting $\Delta m$ from Eq.\ (\ref{dm}) and the center of gravity $m$ from Eq.\ (\ref{m}) as a function of nuclear matter density $n$. The colored region indicates the confidence range.}
\label{fig2}
\end{figure}

QCD sum rules offer the unique possibility to relate (non-perturbative) QCD parameters and hadronic observables. To be specific, the current-current correlation function
\begin{eqnarray} \label{eq1}
\Pi(q) &=
\rmi \int \rmd^4x \, \rme^{\rmi qx} \langle \langle \, T \left[ j(x) j\,^\dagger(0) \right] \rangle \rangle \: ,
\end{eqnarray}
with currents $j(x)$ (with the quark structure $\rmi\bar{d}(x) \gamma_5c(x)$ for the pseudo-scalar $D$ mesons, for instance) is related to the integrated spectral density
\begin{eqnarray}
\Pi(q) &=
\frac{1}{\pi} \int_{-\infty}^{+\infty} \rmd s
\frac{\Delta \Pi(s,|\vec{q}\,|)}{s-q_0} \: ,
\end{eqnarray}
where $\Delta \Pi$ in turn is related to hadronic observables.
We are here interested in an exploration of spectral changes.
Quite general, one may consider the moments
\begin{eqnarray}
\label{mom}
S_n(M) \equiv
\int_{s_0^-}^{s_0^+} \rmd s \, s^n
\Delta \Pi(s) \rme^{-s^2/M^2}
\end{eqnarray}
which allow to define the quantities
\begin{eqnarray} 
{\Delta m} & \equiv
\frac12 \frac{ S_1 S_2 - S_0 S_3}{S_1^2 - S_0 S_2} \: ,
\label{dm}
\\
{ m_+ m _-} & \equiv
- \frac{S_2^2 - S_1 S_3}{S_1^2 - S_0 S_2} \: ,
\\
{m}^2 & \equiv { \Delta m }^2 + {m_+ m _-} \: ,
\label{m}
\end{eqnarray}
which may be interpreted, e.g.\ by exploiting the pole + continuum ansatz, as mass splitting (actually, $\Delta m$ is half of the splitting) and mass shift of $D$ and $\bar D$.
In line with a similar pattern in the strange meson sector, one expects shifts of spectral strengths by an ambient nuclear medium as exhibited schematically in figure \ref{fig1}.
An evaluation of the above equations leads to the results displayed in figure \ref{fig2} (for details and a discussion of different parameter sets cf. \cite{hilger,hilger2}).
These results point to a negative mass splitting $D^+-D^-$ (left panel), while a safe average shift of $D^+ + D^-$ cannot be extracted from the present approach due to uncertainties of input parameters (see right panel of figure \ref{fig2}).
Similar considerations apply for $B$ and $D_s$ mesons \cite{hilger}.

\begin{figure}[ht]
\centering
\includegraphics[width=0.45\linewidth]{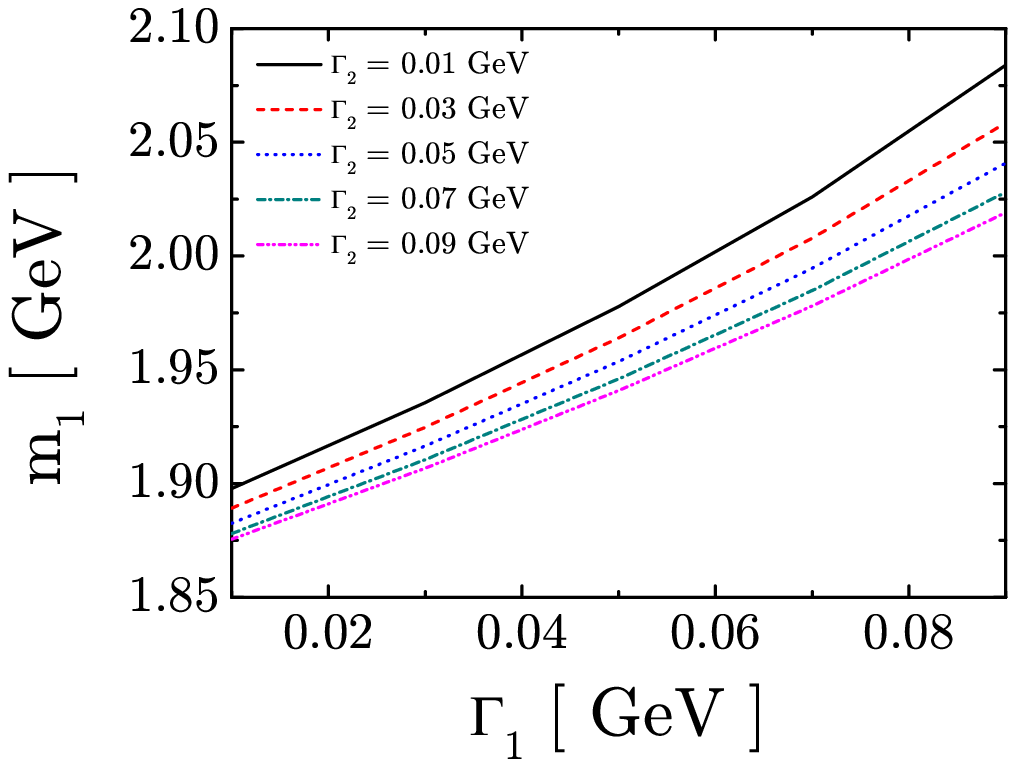}
\includegraphics[width=0.45\linewidth]{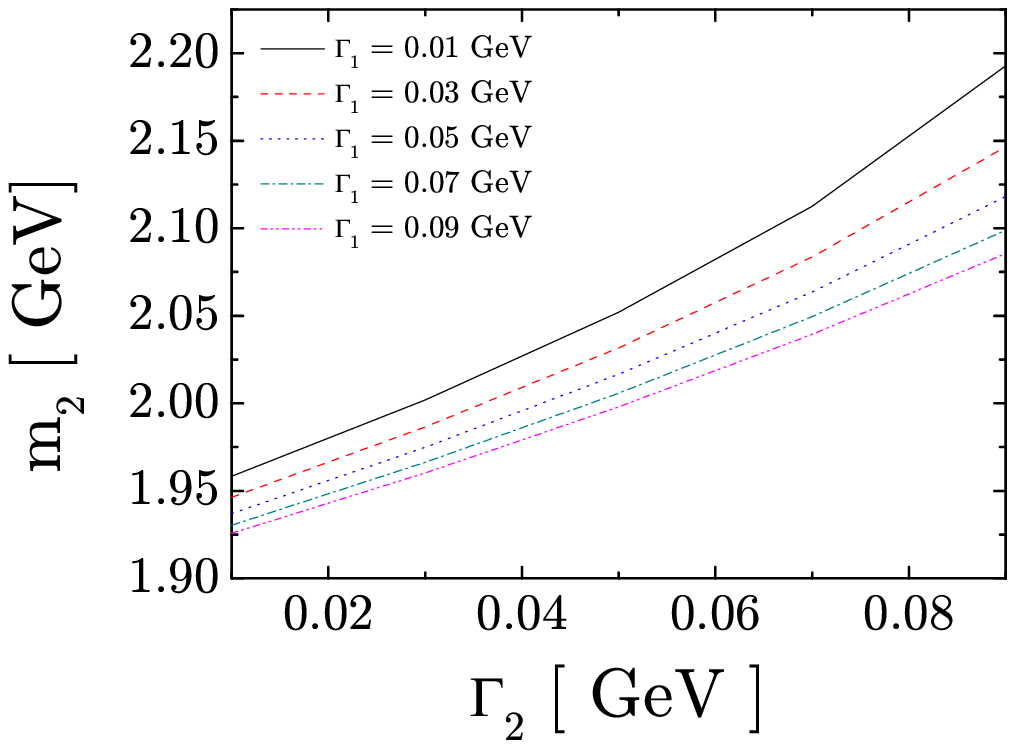}
\caption{Mass--width correlations for $m_+$ (left panel) and $m_-$ (right panel) for different assumptions of widths of the respective other particle.
$\Delta m = -30$ MeV, $m = 1.915$ GeV and $F_+ = F_-$ are assumed.
In a first step, the integration limits in Eq.\ (\ref{mom}) have been extended to $\pm \infty$ without including the continuum contribution, to avoid the influence of possible threshold effects.}
\label{figGm}
\end{figure}

A discussion of finite width effects always meets the problem of additional parameters which have to be fixed by the same set of equations.
Hence, one only is able to give the mass of the particle (i.e.\ the center of gravity of the spectral function) as a function of its width.
In the case of $D$ mesons the widths for particle ($+$) and anti-particle ($-$) enter, and the mass--width correlation of the particle is locked with the width of the antiparticle as well.
Employing the following Breit-Wigner ansatz for the spectral function
\begin{equation}
	\Delta \Pi(\omega) = \frac{F_+}{\pi} \frac{ \omega \Gamma_+}{ (\omega^2-m_+^2)^2 + \omega^2 \Gamma_+^2 } \Theta(\omega)
		+ \frac{F_-}{\pi} \frac{ \omega \Gamma_-}{ (\omega^2-m_-^2)^2 + \omega^2 \Gamma_-^2 } \Theta(-\omega)
\end{equation}
and determining $m_\pm$, $\Gamma_\pm$ such that they fulfill Eqs.\ (\ref{dm}) and (\ref{m}). For $\Delta m = -30$ MeV and $m = 1.915$ GeV one obtains the results depicted in figure \ref{figGm}.

\section{Weinberg type sum rules}
\label{sct3}

Exploiting the Operator Product Expansion one can evaluate in the spirit of (\ref{eq1}) the differences of spectral moments for vector and axial-vector currents, as has been pioneered by Weinberg \cite{Weinberg}, and applied to a hot medium by Shuryak and Kapusta \cite{Kapusta}.
Our results for the differences of vector (V) -- axial-vector (A) and the scalar (S) -- pseudo-scalar (P) correlators in case of mesons with one light quark, $m_1 \ll \Lambda_{QCD}$, and one heavy quark, $m_Q = m_2 \gg \Lambda_{QCD}$, are summarized in table \ref{tbl1}.
\begin{table}
\begin{center}
	\begin{tabular}{cccc}
	\br
		l & $V-A$, light-light & $V-A$, heavy-light & $P-S$, heavy-light
		\\
		\mr
		$-1$ & $F_\pi^2$ & &
		\\
		\hline
		$0$ & $0$ & $8m_Q\langle\bar{q} q\rangle$ & $2 m_Q\langle\bar{q} q\rangle$
		\\
		\hline
		$1$
		&
		$ \mathcal{O}_6 $
		&
		$8 m_Q^3 \langle \overline{q} q \rangle + 4m_Q \langle \Delta \rangle$
		&
		\parbox[]{4cm}{\center $- 2 m_Q^3 \langle \overline{q} q \rangle + m_Q \langle \overline{q} g \sigma \mathcal{G} q \rangle - m_Q \langle \Delta \rangle$}
		\\
		\hline
		$2$
		&
		&
		\parbox{4cm}{\center $8 m_Q^5 \langle \overline{q} q \rangle -4m_Q^3 \langle \overline{q} g \sigma \mathcal{G} q \rangle - 12 \langle \Delta \rangle + \mathcal{O}_7$}
		&
		\parbox{4cm}{\center $- 2 m_Q^5 \langle \overline{q} q \rangle + 3 m_Q^3 \langle \overline{q} g \sigma \mathcal{G} q \rangle - 3 \langle \Delta \rangle + \mathcal{O}_7$}
		\\
		\br
	\end{tabular}
	\caption{A few selected moments of the spectral differences $\int \rmd s s^l \left( \Delta \Pi_{V(P)} - \Delta \Pi_{A(S)} \right)$ for chiral partners.
	$\mathcal{O}_i$ denotes a condensate or a combination of condensates of mass dimension $i$ and $\langle \Delta \rangle$ stands for a medium-specific combination of condensates, i.e.\ it vanishes at zero density and temperature.}
	\label{tbl1}
\end{center}
\end{table}
The employed currents are
$j_\mu^{\mathrm{V(A)}} = \overline{q}_1 (\gamma_5) \gamma_\mu q_2$,
$j^{\mathrm{S(P)}} = (\rmi) \overline{q}_1 (\gamma_5) q_2$.
In table \ref{tbl1}, also the first spectral difference moments for mesons consisting of two light quarks, $m_{1,2} \ll \Lambda_{QCD}$ are listed for $V-A$.

In particular, one recognizes for heavy-light mesons that the differences between chiral partners are driven by the chiral condensate $\langle \bar{q} q \rangle$ in conjunction with the heavy quark mass $m_Q$.
Furthermore, in case of light-light vector mesons the first condensate which qualifies as an order parameter is a certain combination of four-quark condensates for the $1$st moment in next-to-leading order of the strong coupling $\alpha_s$, whereas the chiral condensate can only occur in lowest order of $\alpha_s$ and in the $0$th moment of the spectral differences if one of the quarks is heavy.
In this respect, one may identify such heavy-light quark bound states (hadrons) as useful probes for chiral restoration to be linked with the density dependence of the chiral condensate $\langle \bar{q} q \rangle_n$.

\section{Condensates near $\bf T_c$}
\label{sct4}

\begin{figure}[ht]
\centering
\includegraphics[width=0.45\textwidth]{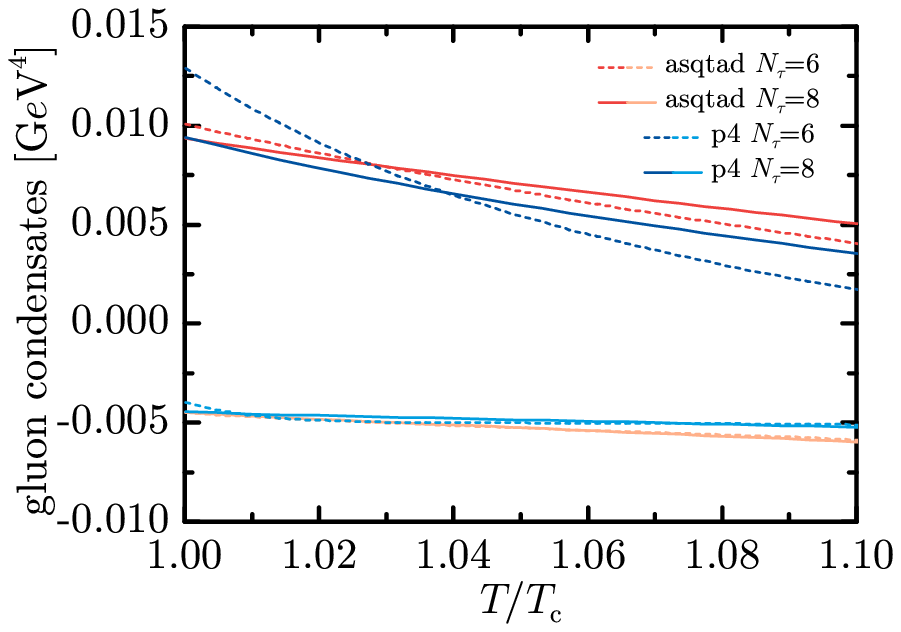}
\includegraphics[width=0.45\textwidth]{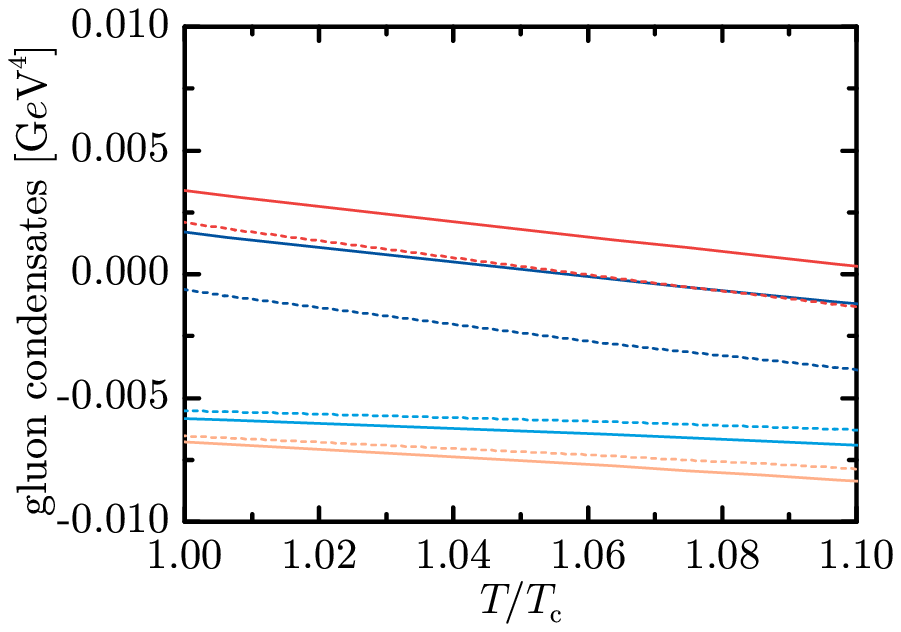}
\caption{Temperature dependence of the gluon condensates (\ref{GGvac}) (upper curves) and (\ref{GGmed}) (lower curves) at $\mu_q = 0$ (left panel) and $\mu_q = 270$ MeV (right panel, same line and color code as in left panel).}
\label{fig3}
\end{figure}

According to the consideration in \cite{SuHongLee} one may relate the gluon condensate at finite temperature to the QCD trace-anomaly for $N_f = 3$ and the equation of state:
\begin{eqnarray}
\left\langle \frac{\alpha_s}{\pi} G^2 \right\rangle_T &=
\left\langle \frac{\alpha_s}{\pi} G^2 \right\rangle_0
- \frac{8}{9} ( e-3p) \: ,
\label{GGvac}
\\
\left\langle \frac{\alpha_s}{\pi} \left( (uG)^2 - \frac{G^2}{4} \right) \right\rangle_T &=
- \frac{3}{4}  \frac{\alpha_s}{\pi} ( e+p ) \: ,
\label{GGmed}
\end{eqnarray}
whereas contributions from light quarks to (\ref{GGvac}) have been omitted in a first step, as we focus on the continuation to finite densities.
The quantities $e-3p$ and $e+p$ are accessible by QCD lattice evaluations \cite{bazavov} and may be extrapolated to finite baryon densities by employing the Rossendorf quasiparticle model \cite{robert} which is here adjusted to \cite{bazavov}.
Both condensates (\ref{GGvac}) and (\ref{GGmed}) are depicted in figure \ref{fig3} in the region near $T_c$ at quark chemical potential $\mu_q = 0$ (left panel) and $\mu_q = 270$ MeV (right panel); energy density $e$ and pressure $p$ have been employed from the quasiparticle model which allows for a thermodynamic consistent extrapolation to finite densities.
The curves for the condensate (\ref{GGvac}) are flattened with increasing lattice temporal extend $N_\tau$; \cite{SuHongLee} is reproduced by the p4 action for $N_\tau = 6$.
On the other hand, the condensate (\ref{GGmed}) seems not to be affected by such a choice.
At finite densities, the gluon condensate for $\langle \frac{\alpha_s}{\pi} G^2 \rangle_T$ drops significantly due to the non-vanishing chemical potential.
The symmetric and traceless gluon condensate $\left\langle \frac{\alpha_s}{\pi} \left( (uG)^2 - \frac{G^2}{4} \right) \right\rangle_T$ is much less influenced by density effects.
These effects have to be studied in detail for the $J/\psi$, which depends essentially on the considered gluon condensates \cite{SuHongLee}.

As mentioned above, finite current quark masses modify the trace of the energy momentum tensor to \cite{Furnstahl} ($N_f = 3$)
\begin{equation}
	T_\mu^\mu = - \frac{9\alpha_s}{8\pi} G_{\mu\nu}^a G^{\mu\nu}_a + m_u \bar{u} u + m_d \bar{d} d + m_s \bar{s} s + \ldots \: ,
\end{equation}
where higher order $\alpha_s$ terms have been neglected.
(The contributions of heavy quarks can be absorbed into the first term by using the heavy-quark mass-expansion.)
A precise analysis would hence also have to include the density and temperature dependence of the quark condensates.
However, as the chiral condensate is supposed to vanish above $T_c$ due to the restoration of chiral symmetry, it is justified to neglect their influence on the gluon condensates in (\ref{GGvac}) and (\ref{GGmed}).
Furthermore, if light quark contributions can be neglected in specific QCD sum rule evaluations, it is consistent with neglecting such terms in (\ref{GGvac}) and (\ref{GGmed}), too.
This offers the avenue towards the evaluation of QCD sum rules near/above the deconfinement transition at non-zero baryon densities.

\section{Summary}

In summary we analyzed in-medium modifications of $D$ mesons in cold nuclear matter and find good prospects for an experimental verification, as the spectral strengths of $D$ and $\bar{D}$ may be displaced by about 60 MeV at saturation density.
The key here is the amplification of the chiral condensate by the heavy charm-quark mass.
We also had a glimpse on an extension of the celebrated Weinberg and Kapusta-Shuryak sum rules for chiral partners and considered, furthermore, the determination of the gluon condensate via the QCD trace anomaly in the deconfinement region at non-zero baryon density.

\section*{Acknowledgments}
The work is supported by BMBF Grant 06DR9059D and GSI-FE.

\end{document}